\documentclass[preprint,12pt]{elsarticle}

\usepackage{graphicx}
\usepackage{subcaption}

\usepackage{amssymb}
\usepackage{amsmath}

\usepackage{lineno}
\usepackage{xcolor}

\newcommand{\captionv}[3]{\begin{center}\parbox{#1cm}{\caption[#2]{{\sf #3}}}
        \end{center}}
\journal{}

\begin{document}
\begin{frontmatter}



\title{Online Dose Verification in VHEE Radiotherapy Using Bremsstrahlung Radiation} 


\author[a,b]{Francesco Urso} 
\author[a,b]{Pietro Carra} 
\author[a,b,c]{Esther Ciarrocchi}
\author[a,b,c]{Matteo Morrocchi}
\author[a,b,c]{Maria Giuseppina Bisogni}

\affiliation[a]{organization={Università di Pisa, Dipartimento di Fisica E. Fermi},%
            addressline={Largo Bruno Pontecorvo 3}, 
            city={Pisa},
            country={Italy}}

\affiliation[b]{organization={Istituto Nazionale di Fisica Nucleare, INFN},%
            addressline={Largo Bruno Pontecorvo 3}, 
            city={Pisa},
            country={Italy}}
\affiliation[c]{organization={Center for Instrument Sharing of the University of Pisa (CISUP), University of Pisa},%
                        city={Pisa},
                        country={Italy}}
            
\begin{abstract}
Very high energy electrons (VHEE) in the 50–250 MeV range, delivered in short pulses at ultra‐high dose rates, are proposed for clinical FLASH radiotherapy (RT) targeting deep-seated tumors. The clinical implementation of VHEE‐FLASH RT requires online verification to optimize dose delivery. In this study we propose a novel online dose verification technique based on the detection of bremsstrahlung photons during VHEE interactions with matter.
A polymethyl methacrylate (PMMA) phantom was simulated to evaluate the dose deposited by a VHEE beam  and to optimise the system design to detect the bremsstralung radiation. Experimental validation was performed at the Beam Test Facility at Laboratori Nazionali di Frascati (Isituto Nazionale di Fisica Nucleare-INFN). A deep learning pipeline was developed to reconstruct the dose distribution in the phantom based on the bremsstrahlung radiation profile.
Experimental results demonstrated successful detection of bremsstrahlung radiation emitted orthogonally to the beam axis. The deep learning model achieved accurate dose reconstruction based on the bremsstrahlung radiation profile with a discrepancy of less than 2\% compared to the simulated dose distribution.
This study confirms that bremsstrahlung detection provides a viable online verification for VHEE RT.\end{abstract}



\end{frontmatter}

\section{Introduction}

Radiotherapy (RT) is one of the main pillars of cancer treatment, using ionizing radiation to target and destroy malignant cells \cite{lievens2020provision,borras2016many}. Although modern RT is performed with increasing precision \cite{hubenak2014mechanisms}, the radiation directed at the tumor inevitably passes through healthy tissues surrounding the affected area. Consequently, research efforts in the field and technological advancements are focused on enhancing the efficacy of radiation in eradicating tumors while minimizing damage to adjacent healthy tissues. 

Among the promising advancements in this field, Very High Energy Electron (VHEE) FLASH radiotherapy stands out for its potential to treat deep‐seated tumors \cite{desrosiers2000150}. VHEE beams, typically in the energy range of 50 to 250 MeV, provide excellent tissue penetration, favorable dose distributions, and are relatively insensitive to tissue heterogeneities \cite{lagzda2020influence}, making them well‐suited for FLASH applications \cite{ronga2021back}. In fact, their higher resilience to density variations enables more precise dose conformity, particularly when combined with advanced treatment planning algorithms \cite{muscato2023treatment,sarti2021deep}. Additionally, spatially fractionated radiation therapy using VHEE, as investigated by Fischer et al. \cite{fischer2024spatially}, has demonstrated promising results in delivering heterogeneous dose distributions that could enhance normal tissue sparing. Furthermore, the relatively small lateral penumbra of VHEE beams ensures better dose conformity and organ-at-risk (OAR) sparing, especially for intracranial and thoracic targets \cite{bohlen2024very}. Even at lower energies (100 MeV), VHEE demonstrates acceptable dose distributions and sparing metrics, albeit with reduced conformity \cite{zhang2023treatment,bohlen2024very}. In recent years, the discovery of the FLASH effect has introduced an opportunity to enhance radiotherapy outcomes by mitigating these side effects. By delivering ultra‐high dose rates ($\geq$40 Gy/s within less than 200 ms), FLASH radiotherapy has shown a remarkable ability to reduce normal tissue toxicity while preserving tumor control \cite{wilson2020ultra,favaudon2014ultrahigh}. VHEE is a versatile approach that addresses technological limitations in FLASH radiotherapy, offering better penetration compared to present low-energy Ultra High Dose Rate (UHDR)  LINAC electron beams.

Despite its promising potential, the clinical adoption of VHEE RT faces significant challenges. VHEE beams require precise magnetic control for stable steering and scanning. However, their lower rigidity compared to protons facilitates beam manipulation \cite{Panaino2024}. VHEE accelerators, including RF linacs and laser–plasma systems, face size and cost barriers, hindering clinical adoption. Compact, high–gradient C-band designs show great promise, although further optimization is required \cite{faillace2022perspectives}.

Accurate secondary dosimetry standardization is a significant step in VHEE’s clinical adoption \cite{Panaino2024}. For instance, radiochromic films offer high spatial resolution and minimal energy dependence up to 20 MeV, remaining dose–rate independent up to $1.5 \times 10^{10}$ Gy/s. However, the 24–48 hour processing time hinders real–time monitoring \cite{devic2011radiochromic}. Although ionization chambers are readily available and suitable for online dosimetry, they can suffer from elevated recombination under nanoseconds VHEE pulses and currently lack standard calibration procedures \cite{Panaino2024}. New dosimetry approaches for UHDR low–energy electron beams include silicon, diamond, plastic scintillators and aerogel detectors \cite{huq2001reference,world2004absorbed, morrocchi2025plastic, ravera20243d, ciarrocchi2024plastic}. For VHEE, Bateman et al. \cite{clements2024mini} and Hart et al. \cite{hart2024plastic} proposed online monitoring solutions for ultra–short pulses. Their work at CLEAR (160–200 MeV) was benchmarked against radiochromic films. Bateman et al. \cite{clements2024mini} introduced the Fibre Optic Flash Monitor, which uses silica fibers and a CMOS camera to detect Cherenkov light. The linear response spanned 0.9–57.4 Gy/pulse with no energy dependence, enabling pulse–by–pulse beam monitoring. Hart et al. \cite{hart2024plastic} tested plastic scintillator detectors with sub–mm resolution. These detectors remained linear up to ~$10^{9}$ Gy/s, though recalibration was needed at kGy doses.

Although these advanced dosimetric techniques are critical to realizing the full potential of VHEE RT, clinical translation will ultimately require quality assurance protocols including in vivo verification of the treatment plan. In vivo dosimetry has been developed to provide direct verification of the dose delivered during external–beam radiotherapy and charged–particle therapy. For clinical linear accelerators, technologies such as electronic portal imaging devices (EPIDs) are widely utilized. EPIDs can acquire high–resolution, two–dimensional dose distributions directly from the treatment beam. In combination with advanced image analysis algorithms, they enable the reconstruction of three–dimensional patient dose distributions and have been integrated with adaptive radiotherapy systems \cite{dogan2023use}. 

In charged–particle therapy, specialized techniques have been developed to meet the specific needs of particle range verification. For example, prompt gamma imaging provides in vivo feedback on range deviations by detecting the secondary radiation produced during particle interactions \cite{krimmer2018prompt}. Other monitoring techniques include positron emission tomography (PET) imaging of the radioactivity produced during the irradiation \cite{parodi2024imaging}. These advancements have improved dosimetric evaluations, with research aiming to enhance detector accuracy, spatial resolution, and integration into automated treatment workflows.

Recent progress in deep learning has opened new opportunities for reconstructing in vivo dose distributions. Deep learning models trained on large datasets of clinical imaging and dosimetric measurements have demonstrated the ability to predict patient–specific three–dimensional dose distributions from input data such as EPID images \cite{marini2024deep}. Such techniques hold promise for improving the accuracy and efficiency of dose reconstruction, especially when dealing with anatomical changes or complex treatment geometries \cite{Babier2020}. In proton and ion beam therapy, deep learning has been employed to estimate dose distributions and range uncertainties using data from PET images and Monte Carlo simulation \cite{jiang20223d,hu2020machine}.
A prominent approach in modern generative modeling uses invertible and differentiable transformations to learn complex probability distributions from data. By starting with simple base densities (e.g., Gaussians) and transforming them into target distributions, one can generate new samples that accurately resemble the observed data. These methods have seen extensive use in fields such as high–energy physics where they provide efficient simulations \cite{Vaselli} and in gravitational–wave inference where their flexibility and scalability are especially valuable \cite{wildberger2024flow}.

While online in vivo dose monitoring has been successfully demonstrated with other radiotherapy modalities, a dedicated approach is still needed for VHEE beams. Hence, we developed a novel in vivo dose verification system specifically designed to address the unique requirements of VHEE RT. In this paper we present a proof–of–concept for an online dose monitoring system that capitalizes on detecting bremsstrahlung radiation generated by the  interactions of the VHEE beam with matter. To develop and validate this monitoring technique, we conducted comprehensive Monte Carlo simulations to analyze dose deposition in a polymethyl methacrylate (PMMA) phantom, which informed the system's design. Experimental validation was performed at the Beam Test Facility (BTF) of the Laboratori Nazionali di Frascati (LNF) of the Istituto Nazionale di Fisica Nucleare (INFN), where bremsstrahlung radiation produced in a PMMA phantom by a VHEE beam was successfully detected and measured. 

To reconstruct the dose distribution in the phantom from the bremsst\-rahlung radiation profile, we implemented a comprehensive deep learning solution. 
We utilized a Flow Matching approach~\cite{lipman2024flow}, which represents an advancement in the generative modeling landscape as a specialized variant of diffusion models~\cite{yang2023diffusion}. 

Unlike conventional diffusion techniques, Flow Matching directly learns the vector field connecting data points to noise distributions, creating more efficient trajectories between them. 
This approach substantially reduces computational overhead during training and inference—a critical advantage for potential clinical applications where real-time feedback is essential. 
The model training relied on Monte Carlo simulation data, with subsequent validation performed using the experimental data collected at the Beam Test Facility.

\section{Methods}

\subsection{Overview of the Online Monitoring System based on secondary radiation}
\begin{figure}[ht]
    \begin{center}
   \includegraphics[width=16cm]{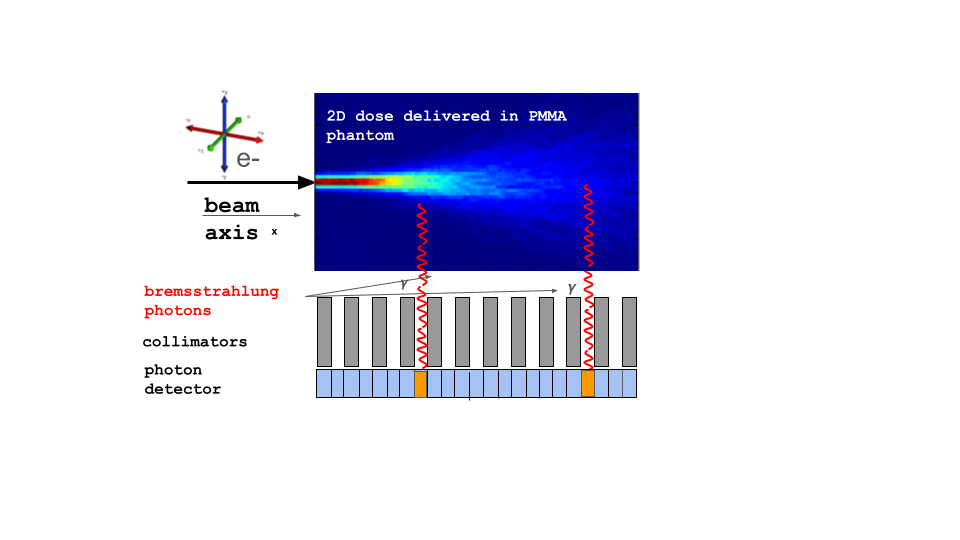}
   \captionv{12}{Short title – can be blank}{Conceptual representation of the secondary radiation online monitoring system for VHEE radiotherapy. The system detects bremsstrahlung photons emitted orthogonally to the beam axis (in red). It samples the radiation profile along the beam axis, producing a histogram (in orange) that can be correlated with the dose distribution within the target area.
   
   \label{schema} 
    }
    \end{center}
\end{figure}

The proposed monitoring system utilizes secondary particles generated during VHEE irradiation. When electrons in this energy range interact with matter, they generate electromagnetic showers, producing bremsstrahlung photons that then create electron–positron pairs primarily via pair production. The system samples the secondary radiation profile along the beam axis by detecting photons emitted orthogonally to the beam direction using a collimator, the majority of which are bremsstrahlung photons with a minor contribution from photons produced during annihilation of the electron-positron pairs. The system selects bremsstrahlung photons emitted at a 90° angle to the beam axis to focus on a specific spatial region, allowing for precise sampling of the area. The resulting radiation profile can be correlated with the absorbed dose through a deep generative algorithm (see Section \ref{sec:deepgenerativemodeling}). Figure \ref{schema} illustrates the conceptual design of this bremsstrahlung-based monitoring system.

\subsection{Monte Carlo Simulation Framework}
Monte Carlo simulations were performed using the Geant4 toolkit \cite{Agostinelli2003,apostolakis2015progress} to model VHEE interactions within a phantom. PMMA, was selected as phantom material for its tissue-equivalent properties, which make it suitable for mimicking human tissue in radiotherapy. Geant4 physics lists were tailored to accurately represent electromagnetic interactions of high–energy electrons, secondary photons, and nuclear interactions. The physics lists used are summarized in Table \ref{tab_physics_lists}.
\begin{table}[htbp]
\begin{center}
\captionv{10}{Physics Lists Implementation}{Summary of the physics lists used in the Geant4 simulations.
\label{tab_physics_lists}
\vspace*{2ex}
}
\begin{tabular} {|l|c|}
\hline
\textbf{Physics List} & \textbf{Description} \\
\hline
G4EmStandardPhysics option4 & Electromagnetic interactions \\\hline
G4HadronElasticPhysics & Elastic hadronic physics \\\hline
G4HadronPhysicsQGSP BIC HP & Inelastic physics interactions for protons and neutrons \\\hline
G4IonBinaryCascadePhysics & Inelastic interactions of other ions \\\hline
G4EmExtraPhysics & Photo–nuclear processes \\\hline
G4StoppingPhysics & Stopping power physics \\
\hline
\end{tabular}
\end{center}
\end{table}

The simulation setup consisted of a cylindrical PMMA phantom with a diameter of 120 mm and a length of 300 mm. The phantom was exposed to electron beams with energies of 150 MeV and 80 MeV, which are suitable for VHEE radiotherapy. The beam width defined as the FWHM of the Gaussian beam, was set to 4 mm. The beam was delivered along the x-axis. The center of the phantom was set at the origin of the reference system and the cylinder axis parallel to the x-axis. The beam entered the phantom at -150 mm. The beam exit window was positioned at -250 mm.
\subsection{Simulated Dose and Secondary Radiation Distributions}
The 1D integrated depth dose (IDD) was calculated in the simulation by integrating the energy deposited in the phantom along the y-axis. The 1D IDD was calculated in the \texttt{SteppingAction} class of the simulation by considering both the energy deposited in each step by the electron and the corresponding position of the energy deposition.

\begin{figure}[ht]
    \begin{center}
    \includegraphics[width=10cm]{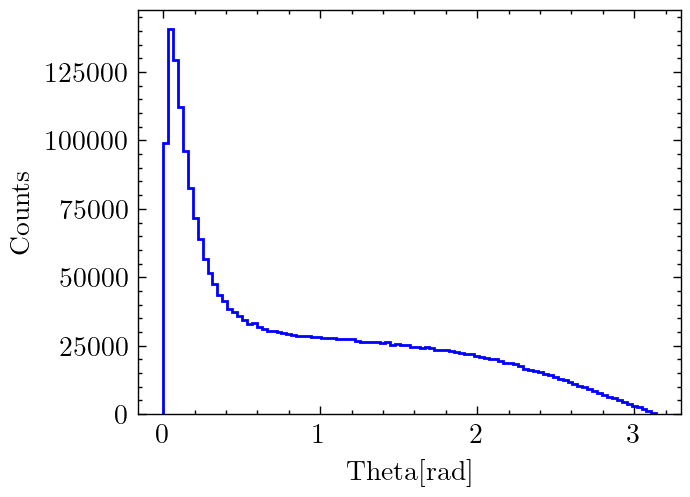}
    \captionv{12}{Short title – can be blank}{Distribution of the emission angles of the photons produced in the phantom for $10^4$ simulated events.
    \label{theta} 
    }
    \end{center}
\end{figure}

\begin{figure}
    \begin{center}
    \includegraphics[width=12cm]{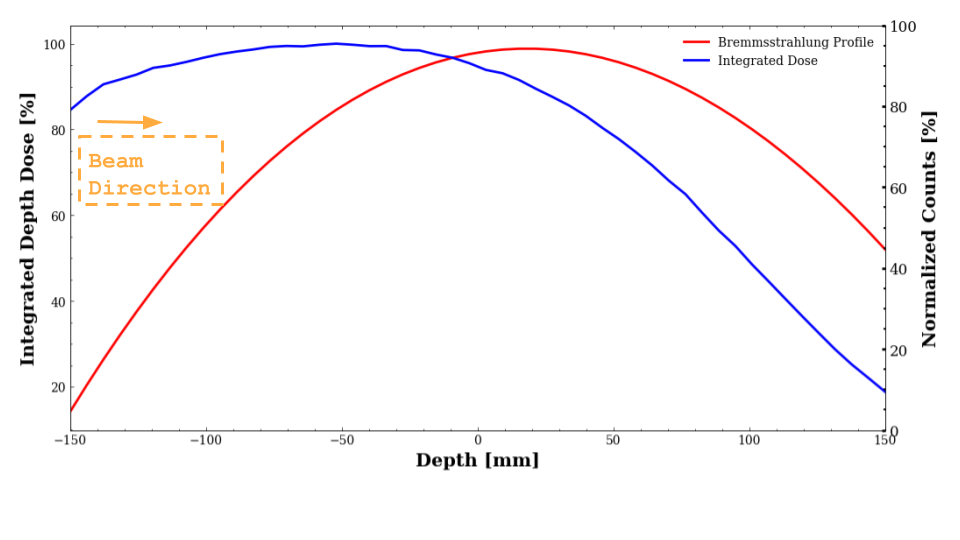}
    \captionv{12}{Short title – can be blank}{Simulated integrated depth dose and number of gamma rays emitted at 90° for 80 MeV electrons interacting in the PMMA phantom for $10^5$ simulated events, both normalized at the maximum value.
    \label{80MeVidd} 
    }
    \end{center}
\end{figure}

If we consider the exiting angle $\theta$ of the photons relative to the direction of the beam, for 10$^4$ primary electrons we obtain the distribution shown in Figure \ref{theta}.
Relativistic electrons produce bremsstrahlung photons predominantly in the direction of the electron’s motion due to the Lorentz boost. To correlate the number of photons detected with a specific region of the phantom, we select only those photons exiting at 90° relative to the phantom’s axis. This selection isolates the region of interest, excluding photons originating from other areas. Figure \ref{80MeVidd} shows for 10$^5$ primary electrons the IDD (blue line) and the number of photons emitted at 90° (red line) for 80 MeV electrons in the simulated phantom both normalised at the maximum value, as a function of the penetration depth in the phantom.

\subsection{Experimental Setup}
The experiments were conducted at the BTF \cite{Buonomo:2023pzi} at LNF. The facility is equipped with a LINAC that produces electron beams with high intensity and low emittance, allowing for precise and controlled experiments. Table \ref{tab_btf_parameters} shows the beam parameters used for the experiments.
\begin{table}[htbp]
\begin{center}
\captionv{10}{Beam Test Facility Parameters}{Summary of the parameters of the beams used at the BTF facility of INFN LNF.
\label{tab_btf_parameters}
\vspace*{2ex}
}
\begin{tabular} {|l|c|}
\hline
\textbf{Particle} & e$^-$ \\\hline
\textbf{Energy (MeV)} & 80/150 \\\hline
\textbf{Repetition rate (Hz)} & 1 \\\hline
\textbf{Pulse length (ns)} & 10 \\\hline
\textbf{Intensity (particle/bunch)} & $10^3$ \\\hline
\hline
\end{tabular}
\end{center}
\end{table}
The experimental setup consisted of a PMMA phantom and two detectors positioned laterally on opposing sides to capture the secondary radiation exiting the phantom. The phantom was composed of three cylindrical sections with a diameter of 12 cm and lengths of 5 cm, 8 cm, and 17 cm, respectively, for a total length of 30 cm.

\begin{figure}[ht]
    \centering
    \begin{subfigure}[b]{0.48\textwidth}
        \centering
        \includegraphics[width=\textwidth]{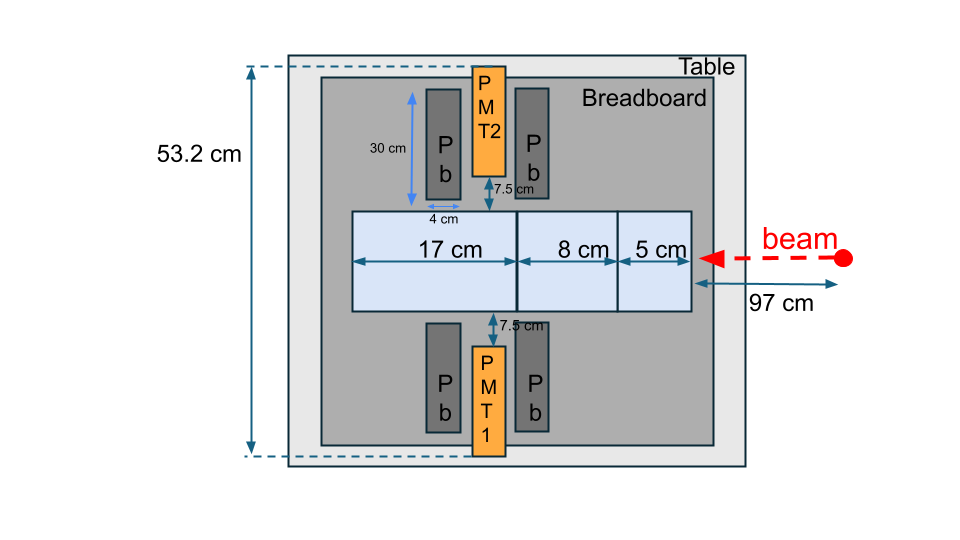}
        \caption{}
        \label{fig:setupschema_a}
    \end{subfigure}%
    \hfill
    \begin{subfigure}[b]{0.48\textwidth}
        \centering
        \includegraphics[width=\textwidth]{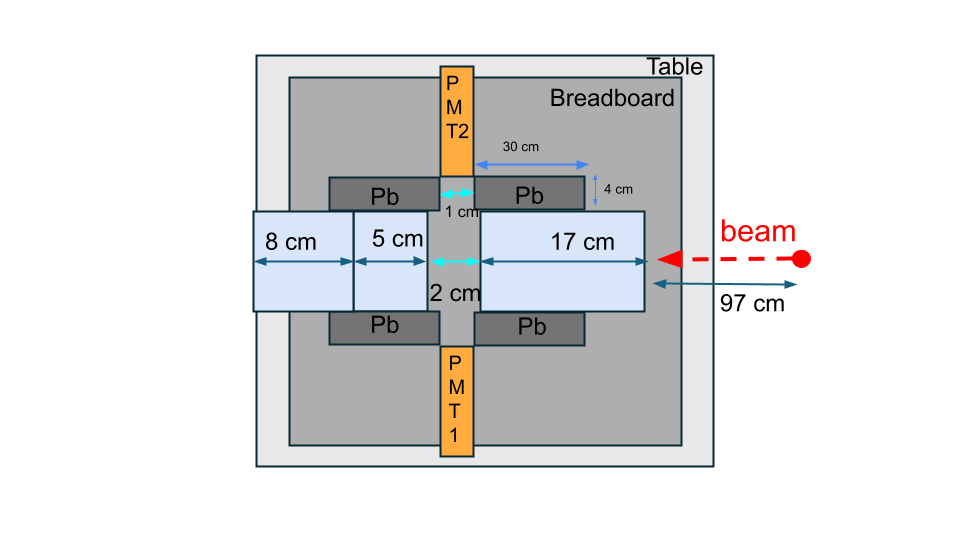}
        \caption{}
        \label{fig:setupschema_b}
    \end{subfigure}

    \vspace{1em}
    \begin{subfigure}[b]{0.48\textwidth}
        \centering
        \includegraphics[width=\textwidth]{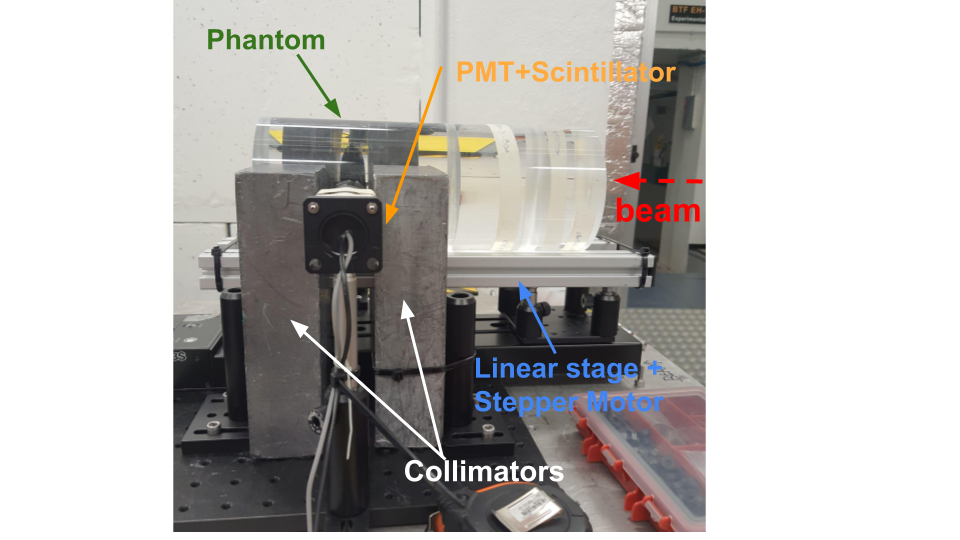}
        \caption{}
        \label{setup_a}
    \end{subfigure}%
    \hfill
    \begin{subfigure}[b]{0.48\textwidth}
        \centering
        \includegraphics[width=\textwidth]{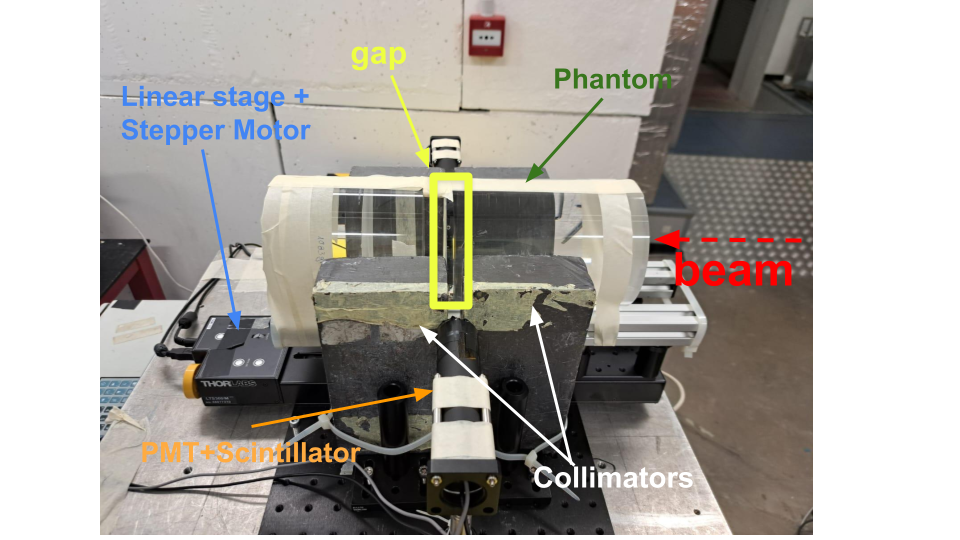}
        \caption{}
        \label{setup_b}
    \end{subfigure}

    \caption{Experimental setup at the BTF at LNF-INFN. (a) Phantom configuration with detectors at 90° to the beam. (b)  With a 2 cm air gap. (c) Photograph of the first setup. (d) Photograph of the setup with a gap in the phantom.}
    \label{fig:setup2x2}
\end{figure}

In the first configuration, illustrated in Figure \ref{fig:setupschema_a}, the PMMA phantom was aligned with the beam axis, and the detectors were placed 7.5 cm from the phantom at 90° to the beam direction. The beam width was measured to be 4 mm using a pixel detector named FitPix \cite{buonomo2016hardware}. Two LYSO crystals (dimensions 1.2×1.2×1 cm$^3$) were coupled to Hamamatsu R9800 Photo Multiplier Tubes (PMTs). The PMTs were connected to an oscilloscope (Lecroy HDO 9204-MS) for signal acquisition and triggering. Lead (Pb) blocks of dimensions 4x14x30 cm$^3$ were used as collimators, as shown in Figure \ref{fig:setupschema_a}.

The phantom was mounted on a linear stage and moved along the beam axis using a stepper motor with a step size of 5 cm, starting at 2.5 cm from the phantom end opposite to the beam entrance. Longitudinal scans of the phantom were performed at energies of 150 MeV and 80 MeV. A photograph of the experimental setup is shown in Figure \ref{setup_a}.
\vspace{1cm}

In the second experimental set up, two sections of the phantom were separated to create an air gap, as shown in Figure \ref{fig:setupschema_b}. The gap was created by rearranging the phantom pieces to leave a 2 cm gap located 17 cm from the phantom's entrance. This setup aimed to investigate the system's ability to detect variations in the bremsstrahlung radiation profile due to the discontinuities in the phantom material.
A photograph of the experimental setup is shown in Figure \ref{setup_b}.
The phantom was irradiated only with the 150 meV beam and the scan was performed with smaller spatial intervals between acquisitions to better capture the profile in the air gap area. The scan in the gap area was performed with a step size of 0.5 cm.

\subsection{Deep Generative Modeling Framework}
\label{sec:deepgenerativemodeling}

The model used was a Continuous Normalizing Flow trained in a Flow-Matching regime\cite{lipman2023flowmatchinggenerativemodeling} \cite{tong2024improvinggeneralizingflowbasedgenerative} described in details in the references. Details on the training configuration and hyperparameters are provided in the \ref{sec:cnfs} of this manuscript.

The model was trained using data from the Geant4 simulation, with the same set-ups described above. The ultimate goal of the model is to reconstruct the dose distribution in the phantom based on the bremsstrahlung radiation profile. To simplify the problem, the data was reduced to one dimension. Thus, the model reconstructs the integrated dose distribution in the phantom based on the integrated bremsstrahlung signal. In practice, this is achieved by conditioning the generation of the dose distribution on the bremsstrahlung signal. Figure \ref{80MeVidd} shows one of the simulated integral 1D bremsstrahlung profile (red line) used for training, as well as the corresponding 1D IDD (blue line) for 80 MeV electrons.

The dataset consisted of simulations for a PMMA cylindrical phantom with a voxel size of 1×1×1\,cm\(^3\) and \(10^{6}\) events. Data were generated for 200 different energies (ranging from 50 to 250\,MeV in steps of 1\,MeV), resulting in a total of 200 different samples of the bremsstrahlung radiation profile and the dose distribution. The model was validated using a separate dataset of 50 samples at energies not present in the training dataset. The model was trained for 1000 epochs.

\section{Results}

\subsection{Experimental Results}

\begin{figure}[ht]
    \begin{center}
    \includegraphics[width=12cm]{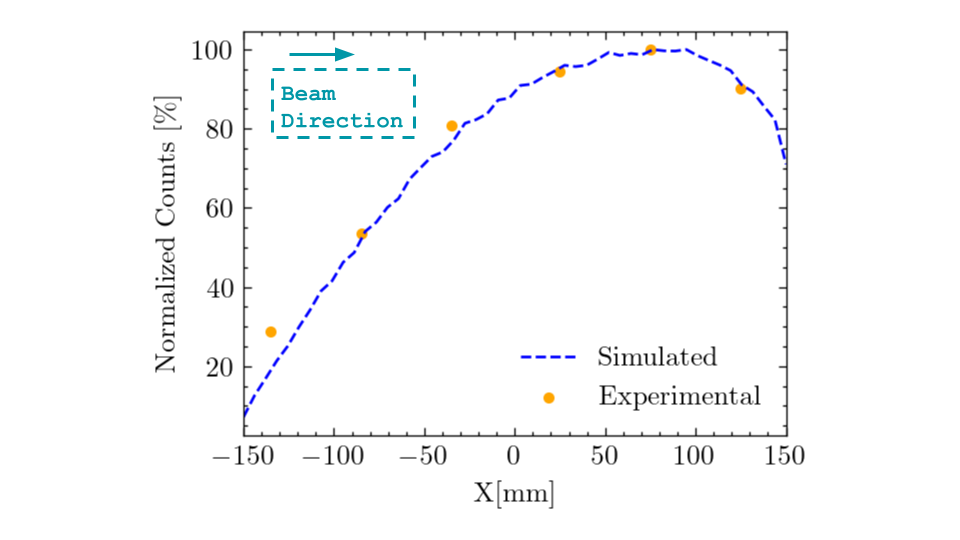}
    \captionv{12}{Short title – can be blank}{Normalized counts acquired during the 150 MeV electron beam scan. The experimental data (orange points) shows the detector counts at different positions along the phantom, while the blue dashed line represents the corresponding simulated bremsstrahlung profile.
    \label{150MeV} 
    }
    \end{center}
\end{figure}

\begin{figure}[ht]
    \begin{center}
    \includegraphics[width=12cm]{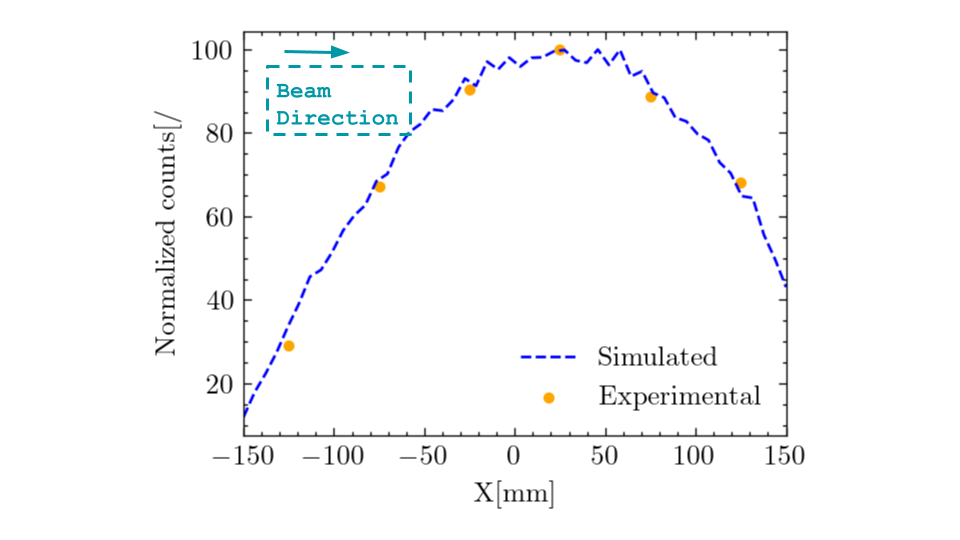}
    \captionv{12}{Short title – can be blank}{Normalized counts acquired during the 80 MeV electron beam scan. The experimental measurements (orange points) demonstrate the characteristic distribution of bremsstrahlung photons detected at 90° to the beam axis, with the blue dashed line showing the simulated profile for comparison. 
    \label{80MeV} 
    }
    \end{center}
\end{figure}

For a given phantom position during the scan, the counts acquired by the detectors were measured and normalized to the maximum value observed during the scan. The statistical uncertainties were calculated as the square root of the number of counts, and the relative errors were propagated accordingly. 
The results are shown in Figure \ref{150MeV} for the 150 MeV scan and in Figure \ref{80MeV} for the 80 MeV scan, respectively. In both cases, the simulated bremsstrahlung profile (blue dashed line) is plotted with the experimental data (orange dots).

\begin{figure}[ht]
    \begin{center}
    \includegraphics[width=10cm]{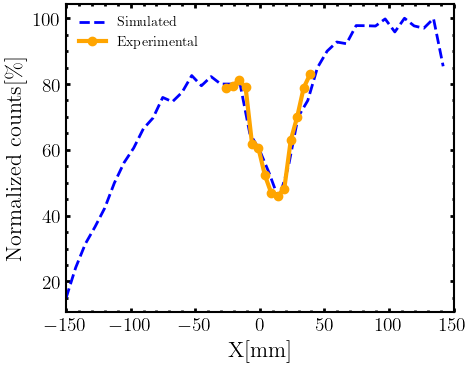}
    \captionv{12}{Short title – can be blank}{Normalized counts acquired during the scan with a 2 cm air gap inserted 17 cm from the phantom entrance with 150 MeV electrons. The experimental data (orange points) clearly reveals the structural discontinuity as a distinctive dip in the profile, while the blue dashed line represents the simulated profile.
    \label{gap} 
    }
    \end{center}
\end{figure}

For the second setup, the counts were normalized to 90 \% of the maximum value observed during the 150 MeV scan in the first setup, to account for the fact that only a fraction of the profile was acquired in the gap scan. The results from the scan in for the second experimental setup are shown in Figure \ref{gap}, the experimental data is represented by the orange points, while the blue dashed line represents the simulated bremsstrahlung profile.

\subsection{Dose Reconstruction}

\begin{figure}[htbp]
    \centering
    \includegraphics[width=0.5\textwidth]{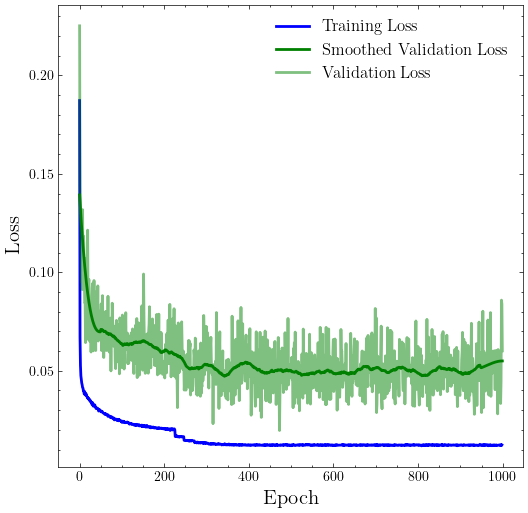}
    \caption{Training and validation loss of the model. The training loss (blue line) and validation loss (green line) are shown as a function of the number of epochs. The validation loss has been smoothed to emphasize the overall trend.} 
    \label{fig:loss}
\end{figure}

\begin{figure}
    \centering
    \includegraphics[width=0.5\textwidth]{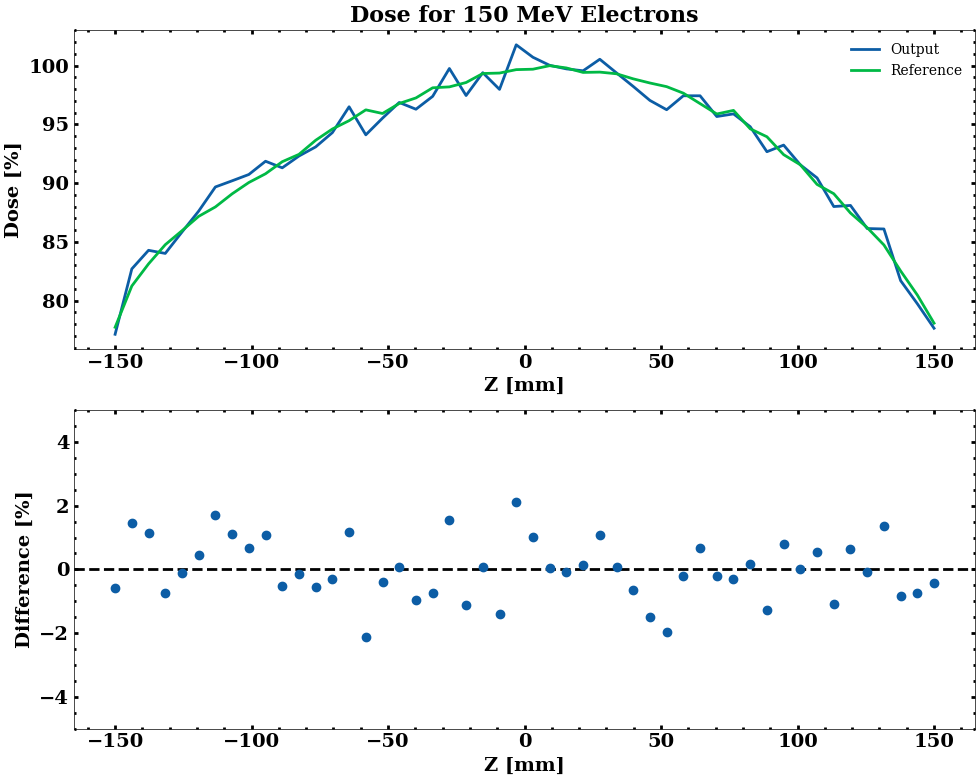}
    \caption{Reconstructed dose distribution for 150 MeV electron beam irradiation. The blue line represents the dose profile reconstructed from experimental bremsstrahlung measurements using the  model, while the green line indicates the ground truth dose distribution from Monte Carlo simulation. The residuals plot (bottom) shows the percentage difference between the reconstructed and true dose profiles.
    }
    \label{150}
\end{figure}

\begin{figure}
    \centering
    \includegraphics[width=0.5\textwidth]{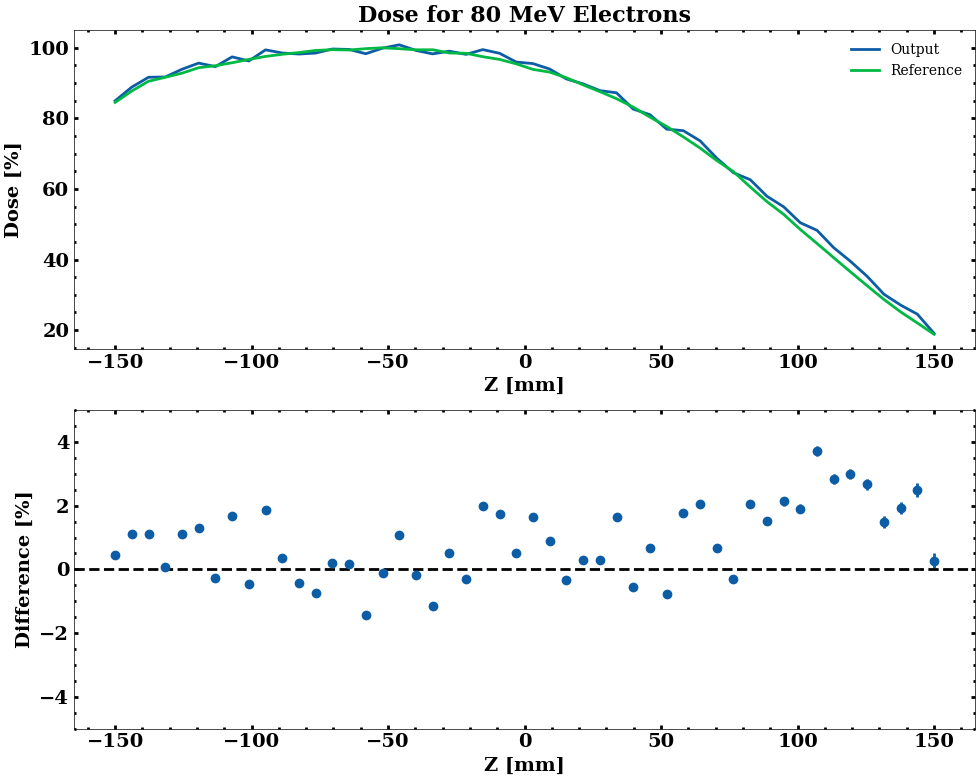}
    \caption{Reconstructed dose distribution for 80 MeV electron beam irradiation. The blue line shows the dose profile reconstructed by the model based on experimental measurements (blue line), compared to the ground truth dose distribution (green line) obtained from simulation.  The residuals plot (bottom) illustrates the percentage difference between the reconstructed and true dose profiles.
    }
    \label{80}
\end{figure}

The experimental data were binned and used to reconstruct the dose distribution in the phantom. The dose was reconstructed using the model described in section \ref{sec:deepgenerativemodeling}.
Figure \ref{fig:loss} shows the training and validation loss as a function of the number of epochs. The training loss (blue line) and validation loss (green line) are shown as a function of the number of epochs. The validation loss has been smoothed to emphasize the overall trend. Owing to the smaller size of the validation dataset, its loss exhibits more variability compared to the training loss. The training loss steadily decreases while the validation loss remains stable, indicating minimal overfitting and satisfactory generalization. More details on the structure of the loss function can be found in \ref{sec:cnfs}.
Figures \ref{150} and \ref{80} show the reconstructed (blue) and true (green) dose distributions for the 150 MeV and 80 MeV scans, respectively, where the ground truth being the simulated dose. 

\section{Discussion}

This study investigates the potential of using a bremsstrahlung-based system for dose verification in VHEE radiotherapy. The profile of the counts acquired during the 150 MeV scan and the 80 MeV, as shown in Figures \ref{150MeV} and \ref{80MeV} respectively, exhibit good agreement. The reduced penetration depth of the 80 MeV electrons is evident in Figure \ref{80MeV}, as the bremsstrahlung profile rapidly decreases in counts due to energy loss in the phantom. The 150 MeV scan, shown in Figure \ref{150MeV}, exhibits a more extended profile, consistent with the higher energy of the electrons. The discrepancy between data and simulation observed at -125 mm in figure \ref{150MeV} could be attributed to:
\begin{itemize}
    \item Higher uncertainties associated with shower development at the tails of the bremsstrahlung profile. 90 degrees exiting photons are in fact produced by electrons of lower energy that have undergone multiple scattering, leading to a wider spread of their distribution in the phantom.
    \item The collimators used in the experimental setup may not have perfectly filtered out all photons emitted at angles other than 90 degrees, leading to a background signal that affects the measurement.
\end{itemize}

The data in Figure \ref{gap} demonstrates the system's ability to identify structural discontinuities within the phantom, with the bremsstrahlung profile clearly highlighting the 2 cm gap. This sensitivity to phantom variations suggests the potential of the method for identifying anatomical changes during treatment.

The dose reconstruction results (Figures \ref{150} and \ref{80}) also underline the effectiveness of the deep generative model: the reconstructed dose distributions successfully replicate the overall shape of the true simulated dose profiles and the discrepancies are within 2\% of the value as shown in the residual plot.
The observed systematic trend in the residuals is small (well below 2\%), and does not significantly affect the accuracy of the reconstructed dose distribution. While further refinement of the model or the inclusion of additional experimental data could help to reduce this residual trend, its current magnitude is unlikely to impact the practical utility of the method. Future work may investigate this effect in more detail, but for the present proof-of-concept, the results demonstrate that the approach is robust and accurate.

\section{Conclusion}
A bremsstrahlung-based online dose monitoring system has been developed as a proof of concept for VHEE radiotherapy. By using Geant4 Monte Carlo simulations with experimental measurements at the LNF Beam Test Facility, we demonstrated that bremsstrahlung photons emitted orthogonally to the beam axis allow to successfully reconstruct the dose distribution in a PMMA phantom.
A Continuous Normalizing Flow–based deep generative model was trained to map the measured bremsstrahlung profiles to integrated depth–dose curves, achieving reconstruction errors between 2\% and 4\%. The system demonstrated the ability to identify structural discontinuities, highlighting its sensitivity to anatomical changes.

The current experimental setup was conducted in a controlled environment with few sources of noise, which is not representative of a clinical beamline.

Looking ahead, the plan is to expand the deep-learning pipeline to reconstruct complete three-dimensional dose volumes using bremsstrahlung measurements. Additionally, a combined simulation–experimental training dataset will be developed to improve model robustness in realistic noise and background conditions. Finally, the integrated system will be tested in a setting that closely mimics clinical practice, incorporating a realistic anthropomorphic phantom.

\section{Acknowledgments}
Funded by European Union  Next Generation EU, Mission 4 Component 1 – “Fund for the National Research Program and for Projects of National Interest (NRP)” Italian Ministry of Education call PRIN 2022 D.D. n. 104 02/02/2022 - Title of the project “Monitor for flash therapy (MORSE)” – ERC PE2-3, project number 20223WS25W-001 CUP I53D23000830006 
\clearpage
\appendix
\section{Model Description and Training Configuration}
\label{app1}

\subsection{Continuous Flows for Generative Modeling}
    
The architecture of the model comprises a ResNet that learns the vector field of the flow. The model was constructed in PyTorch \cite{lipman2023flowmatchinggenerativemodeling}, using the \texttt{torchdiffeq} package for the ODE solver and \texttt{TorchCFM} for building the Conditional Flow Matching model (CFM) \cite{tong2024improvinggeneralizingflowbasedgenerative}.

\label{sec:cnfs}

To transform an initial distribution into another, one can adopt a continuous–time formulation driven by an ordinary differential equation (ODE). In this formulation, a vector field \(v_t(x)\), parameterized by a neural network, evolves the initial density (e.g., Gaussian) into the target distribution without relying on piecewise compositions. Formally,

\begin{equation}
    \frac{d\phi_t(x)}{dt} = v_t\bigl(\phi_t(x)\bigr),\quad \phi_0(x) = x,
\end{equation}
where \(\phi_t\) maps samples from the base distribution at \(t=0\) to the data distribution at \(t=1\). The invertibility of this procedure ensures that the resulting probability density can be evaluated exactly, enabling likelihood–based training.

The CFM loss aims to train the learned vector field \(v_t(x;\theta)\) to match a target field \(u_t(x\mid x_1)\) along the path from a simple distribution to the data point \(x_1\). In the loss function
\begin{equation}
    L_{CFM}(\theta) = \mathbb{E}_{t,\,q(x_1),\,p_t(x \mid x_1)}\!\bigl[\|v_t(x;\theta) - u_t(x\mid x_1)\|^{2}\bigr],
\end{equation}
the terms are:
- \(v_t(x;\theta)\): the learned vector field at time \(t\) parameterized by \(\theta\),
- \(u_t(x\mid x_1)\): the target vector field that guides the transformation toward the data point \(x_1\),
- \(\mathbb{E}_{t,\,q(x_1),\,p_t(x \mid x_1)}\): the expectation taken over time \(t\), data points \(x_1\) sampled from the distribution \(q(x_1)\), and trajectories \(x\) sampled from the conditional path distribution \(p_t(x \mid x_1)\),
- \(\|\cdot\|^{2}\): the squared Euclidean norm which quantifies the difference between \(v_t(x;\theta)\) and \(u_t(x\mid x_1)\).

\subsection{Training Configuration and Hyperparameters}
Table \ref{tab:hyperparameters} outlines the main hyperparameters of the model. The target and input dimensions are both 50, reflecting the number of features used for dose and bremsstrahlung signals, respectively. The model was trained for 1000 epochs with a 0.001 learning rate, using the Adam optimizer. A \texttt{ReduceLROnPlateau} scheduler adapts the learning rate based on validation performance. The dataset includes 67\,500 training samples and 500 test samples, processed in batches of 100, with Gaussian noise added to replicate realistic conditions. CFM employs a minimum \(\sigma_{\min}\) of 0.0001, with the target–matching objective. The ODE backend is implemented via \texttt{torchdiffeq} using the Euler method, and the \(\alpha\) parameter is set to 1 across 100 timesteps. The ResNet architecture has hidden dimensions \([32\times1,\;48\times2,\;128\times2,\;32\times2]\), uses GELU activation functions, a dropout rate of 0.1, and omits batch normalization, yielding a total of 26,569 parameters.

 \begin{table}[htbp]
     \centering
     \caption{Hyperparameter Values and Configuration}
     \begin{tabular}{|l|l|}
     \hline
     \textbf{Hyperparameter} & \textbf{Value} \\ \hline
     \multicolumn{2}{|c|}{\textbf{Target and Input Dimensions}} \\ \hline
     Target (Dose) Dimension & 50 \\ \hline
     Input (Bremsstrahlung Profile) Dimension & 50 \\ \hline
     \multicolumn{2}{|c|}{\textbf{Training Parameters}} \\ \hline
     Epochs & 1000 \\ \hline
     Learning Rate & 0.001 \\ \hline
     Optimizer & Adam \\ \hline
     Scheduler & ReduceLROnPlateau \\ \hline
     \multicolumn{2}{|c|}{\textbf{Data Parameters}} \\ \hline
     Number of Training Samples & 200 \\ \hline
     Number of Test Samples & 50 \\ \hline
     Batch Size & 10 \\ \hline
     \multicolumn{2}{|c|}{\textbf{Model Parameters}} \\ \hline
     CFM \(\sigma_{min}\) & 0.0001 \\ \hline
     Matching Type & Target \\ \hline
     ODE Backend & torchdiffeq, Euler \\ \hline
     \(\alpha\) & 1 \\ \hline
     Timesteps & 100 \\ \hline
     Type & ResNet \\ \hline
     Hidden Dimensions & [32x1, 48x2, 128x2, 32x2] \\ \hline
     Activation Function & GELU \\ \hline
     Dropout & 0.1 \\ \hline
     Batch Normalization & False \\ \hline
     Total Parameters & 26569 \\ \hline
     \end{tabular}
    \label{tab:hyperparameters}

 \end{table}


\clearpage

\begin{thebibliography}{00}


\bibitem[Lievens et al.(2020)]{lievens2020provision}
Y. Lievens, J. M. Borras, C. Grau. Provision and use of radiotherapy in Europe. Molecular oncology, 14(7):1461--1469, 2020.

\bibitem[Borras et al.(2016)]{borras2016many}
J. M. Borras, Y. Lievens, M. Barton, J. Corral, J. Ferlay, F. Bray, C. Grau. How many new cancer patients in Europe will require radiotherapy by 2025? An ESTRO-HERO analysis. Radiotherapy and Oncology, 119(1):5--11, 2016.

\bibitem[Hubenak et al.(2014)]{hubenak2014mechanisms}
J. R. Hubenak, Q. Zhang, C. D. Branch, S. J. Kronowitz. Mechanisms of injury to normal tissue after radiotherapy: a review. Plastic and reconstructive surgery, 133(1):49e--56e, 2014.


\bibitem[DesRosiers et al.(2000)]{desrosiers2000150}
C. DesRosiers, V. Moskvin, A. F. Bielajew, L. Papiez. 150-250 MeV electron beams in radiation therapy. Physics in Medicine \& Biology, 45(7):1781, 2000.

\bibitem[Lagzda et al.(2020)]{lagzda2020influence}
A. Lagzda, D. Angal-Kalinin, J. Jones, A. Aitkenhead, K. J. Kirkby, R. MacKay, M. Van Herk, W. Farabolini, S. Zeeshan, R. M. Jones. Influence of heterogeneous media on Very High Energy Electron (VHEE) dose penetration and a Monte Carlo-based comparison with existing radiotherapy modalities. Nuclear Instruments and Methods in Physics Research Section B: Beam Interactions with Materials and Atoms, 482:70--81, 2020.

\bibitem[Ronga et al.(2021)]{ronga2021back}
M. G. Ronga, M. Cavallone, A. Patriarca, A. M. Leite, P. Loap, V. Favaudon, G. Cr{\'e}hange, L. De Marzi. Back to the future: very high-energy electrons (VHEEs) and their potential application in radiation therapy. Cancers, 13(19):4942, 2021.

\bibitem[Muscato et al.(2023)]{muscato2023treatment}
A. Muscato, L. Arsini, G. Battistoni, L. Campana, D. Carlotti, F. De Felice, A. De Gregorio, M. De Simoni, C. Di Felice, Y. Dong, et al. Treatment planning of intracranial lesions with VHEE: comparing conventional and FLASH irradiation potential with state-of-the-art photon and proton radiotherapy. Frontiers in Physics, 11:1185598, 2023.

\bibitem[Sarti et al.(2021)]{sarti2021deep}
A. Sarti, P. De Maria, G. Battistoni, M. De Simoni, C. Di Felice, Y. Dong, M. Fischetti, G. Franciosini, M. Marafini, F. Marampon, et al. Deep seated tumour treatments with electrons of high energy delivered at FLASH rates: the example of prostate cancer. Frontiers in Oncology, 11:777852, 2021.

\bibitem[Fischer et al.(2024)]{fischer2024spatially}
J. Fischer, A. J. Hart, N. Bedriova, D. Krim, N. Clements, J. J. Bateman, P. Korysko, W. Farabolini, V. Rieker, R. Corsini, et al. Spatially Fractionated Radiotherapy with Very High Energy Electron Pencil Beam Scanning. Physics in Medicine and Biology, 2024.

\bibitem[B{\"o}hlen et al.(2024)]{bohlen2024very}
T. T. B{\"o}hlen, J. F. Germond, L. Desorgher, I. Veres, A. Bratel, E. Landstr{\"o}m, E. Engwall, F. G. Herrera, E. M. Ozsahin, J. Bourhis, et al. Very high-energy electron therapy as light-particle alternative to transmission proton FLASH therapy--An evaluation of dosimetric performances. Radiotherapy and Oncology, 194:110177, 2024.
\bibitem[Zhang et al.(2023)]{zhang2023treatment}
G. Zhang, Z. Zhang, W. Gao, H. Quan. Treatment planning consideration for very high-energy electron FLASH radiotherapy. Physica Medica, 107:102539, 2023.

\bibitem[Wilson et al.(2020)]{wilson2020ultra}
J. D. Wilson, E. M. Hammond, G. S. Higgins, K. Petersson. Ultra-high dose rate (FLASH) radiotherapy: Silver bullet or fool's gold? Frontiers in oncology, 9:1563, 2020.

\bibitem[Favaudon et al.(2014)]{favaudon2014ultrahigh}
V. Favaudon, L. Caplier, V. Monceau, F. Pouzoulet, M. Sayarath, C. Fouillade, M. F. Poupon, I. Brito, P. Hup{\'e}, J. Bourhis, et al. Ultrahigh dose-rate FLASH irradiation increases the differential response between normal and tumor tissue in mice. Science translational medicine, 6(245):245ra93--245ra93, 2014.

\bibitem[Panaino et al.(2024)]{Panaino2024}
C. M. V. Panaino, et al. Very High-Energy Electron Therapy Toward Clinical Implementation: A Review Study. Preprints, 2024. doi:10.20944/preprints202411.0913.v1.



\bibitem[Faillace et al.(2022)]{faillace2022perspectives}
L. Faillace, D. Alesini, G. Bisogni, F. Bosco, M. Carillo, P. Cirrone, G. Cuttone, D. De Arcangelis, A. De Gregorio, F. Di Martino, et al. Perspectives in linear accelerator for FLASH VHEE: Study of a compact C-band system. Physica Medica, 104:149--159, 2022.

\bibitem[Devic(2011)]{devic2011radiochromic}
S. Devic. Radiochromic film dosimetry: past, present, and future. Physica medica, 27(3):122--134, 2011.

\bibitem[Huq et al.(2001)]{huq2001reference}
M. S. Huq, H. Song, P. Andreo, C. J. Houser. Reference dosimetry in clinical high-energy electron beams: Comparison of the AAPM TG-51 and AAPM TG-21 dosimetry protocols. Medical physics, 28(10):2077--2087, 2001.

\bibitem[WHO(2004)]{world2004absorbed}
World Health Organization. Absorbed dose determination in external beam radiotherapy. An international code of practice for dosimetry based on standards of absorbed dose to water, 2004.
\bibitem[Morrocchi et al.(2025)]{morrocchi2025plastic}
M. Morrocchi, E. Ciarrocchi, R. Anzalone, A. Cavalieri, F. Di Martino, C. D'Orazio, M. Massa, A. Moggi, C. Mozzo, J. H. Pensavalle, et al. Plastic scintillator sheets for quality assurance in electron FLASH and minibeam radiation therapy. Medical Physics, 52(8):e18033, 2025.

\bibitem[Ravera et al.(2024)]{ravera20243d}
E. Ravera, R. Anzalone, A. Cavalieri, E. Ciarrocchi, D. Del Sarto, F. Di Martino, M. Morrocchi, M. G. Bisogni. A 3D imaging system for dosimetry in FLASH radiotherapy. Nuclear Instruments and Methods in Physics Research Section A: Accelerators, Spectrometers, Detectors and Associated Equipment, 1069:169910, 2024.

\bibitem[Ciarrocchi et al.(2024)]{ciarrocchi2024plastic}
E. Ciarrocchi, E. Ravera, A. Cavalieri, M. Celentano, D. Del Sarto, F. Di Martino, S. Linsalata, M. Massa, L. Masturzo, A. Moggi, et al. Plastic scintillator-based dosimeters for ultra-high dose rate (UHDR) electron radiotherapy. Physica Medica, 121:103360, 2024.

\bibitem[Clements et al.(2024)]{clements2024mini}
N. Clements, N. Esplen, J. Bateman, C. Robertson, M. Dosanjh, P. Korysko, W. Farabolini, R. Corsini, M. Bazalova-Carter. Mini-GRID radiotherapy on the CLEAR very-high-energy electron beamline: collimator optimization, film dosimetry, and Monte Carlo simulations. Physics in Medicine \& Biology, 69(5):055003, 2024.

\bibitem[Hart et al.(2024)]{hart2024plastic}
A. Hart, C. Gigu{\`e}re, J. Bateman, P. Korysko, W. Farabolini, V. Rieker, N. Esplen, R. Corsini, M. Dosanjh, L. Beaulieu, et al. Plastic scintillator dosimetry of ultrahigh dose-rate 200 MeV electrons at CLEAR. IEEE Sensors Journal, 2024.

\bibitem[Dogan et al.(2023)]{dogan2023use}
N. Dogan, B. J. Mijnheer, K. Padgett, A. Nalichowski, C. Wu, M. J. Nyflot, A. J. Olch, N. Papanikolaou, J. Shi, S. M. Holmes, et al. Use of electronic portal imaging devices for pre-treatment and in vivo dosimetry patient-specific IMRT and VMAT QA: Report of AAPM Task Group 307. Medical physics, 50(8):e865, 2023.

\bibitem[Krimmer et al.(2018)]{krimmer2018prompt}
J. Krimmer, D. Dauvergne, J. M. L{\'e}tang, {\'E}. Testa. Prompt-gamma monitoring in hadrontherapy: A review. Nuclear Instruments and Methods in Physics Research Section A: Accelerators, Spectrometers, Detectors and Associated Equipment, 878:58--73, 2018.

\bibitem[Parodi(2024)]{parodi2024imaging}
K. Parodi. Imaging for ion beam therapy: current trends and future perspectives. Health and Technology, 1--7, 2024.

\bibitem[Marini et al.(2024)]{marini2024deep}
L. Marini, M. Avanzo, A. C. Kraan, F. Lizzi, C. Mozzi, A. Retico, C. Talamonti. Deep learning methods for 2D in-vivo dose reconstruction with EPID detector. Nuclear Instruments and Methods in Physics Research Section A: Accelerators, Spectrometers, Detectors and Associated Equipment, 1069:169908, 2024.

\bibitem[Babier and Delaney(2020)]{Babier2020}
A. Babier, A. R. Delaney. Deep learning applications in radiation therapy: Dosimetric accuracy and clinical implementation. Medical Physics, 47:e185-e196, 2020.

\bibitem[Jiang et al.(2022)]{jiang20223d}
Z. Jiang, L. Sun, W. Yao, Q. J. Wu, L. Xiang, L. Ren. 3D in vivo dose verification in prostate proton therapy with deep learning-based proton-acoustic imaging. Physics in Medicine \& Biology, 67(21):215012, 2022.

\bibitem[Hu et al.(2020)]{hu2020machine}
Z. Hu, G. Li, X. Zhang, K. Ye, J. Lu, H. Peng. A machine learning framework with anatomical prior for online dose verification using positron emitters and PET in proton therapy. Physics in Medicine \& Biology, 65(18):185003, 2020.

\bibitem[Vaselli et al.(2024)]{Vaselli}
F. Vaselli, F. Cattafesta, P. Asenov, A. Rizzi. End-to-end simulation of particle physics events with Flow Matching and generator Oversampling. Machine Learning: Science and Technology, 2024.

\bibitem[Wildberger et al.(2024)]{wildberger2024flow}
J. Wildberger, M. Dax, S. Buchholz, S. Green, J. H. Macke, B. Sch{\"o}lkopf. Flow matching for scalable simulation-based inference. Advances in Neural Information Processing Systems, 36, 2024.

\bibitem[Tong et al.(2024)]{tong2024improvinggeneralizingflowbasedgenerative}
A. Tong, K. Fatras, N. Malkin, G. Huguet, Y. Zhang, J. Rector-Brooks, G. Wolf, Y. Bengio. Improving and generalizing flow-based generative models with minibatch optimal transport. arXiv preprint arXiv:2302.00482, 2024.

\bibitem[Lipman et al.(2023)]{lipman2023flowmatchinggenerativemodeling}
Y. Lipman, R. T. Q. Chen, H. Ben-Hamu, M. Nickel, M. Le. Flow Matching for Generative Modeling. arXiv preprint arXiv:2210.02747 [cs.LG], 2023. Available at: https://arxiv.org/abs/2210.02747

\bibitem[Lipman et al.(2024)]{lipman2024flow}
Y. Lipman, M. Havasi, P. Holderrieth, N. Shaul, M. Le, B. Karrer, R. T. Q. Chen, D. Lopez-Paz, H. Ben-Hamu, I. Gat. Flow matching guide and code. arXiv preprint arXiv:2412.06264, 2024.

\bibitem[Yang et al.(2023)]{yang2023diffusion}
L. Yang, Z. Zhang, Y. Song, S. Hong, R. Xu, Y. Zhao, W. Zhang, B. Cui, M.-H. Yang. Diffusion models: A comprehensive survey of methods and applications. ACM Computing Surveys, 56(4):1--39, 2023.

\bibitem[Agostinelli et al.(2003)]{Agostinelli2003}
S. Agostinelli, et al. GEANT4: a simulation toolkit. Nuclear instruments and methods in physics research section A: Accelerators, Spectrometers, Detectors and Associated Equipment, 506(3):250--303, 2003.

\bibitem[Apostolakis et al.(2015)]{apostolakis2015progress}
J. Apostolakis, M. Asai, A. Bagulya, J. M. C. Brown, H. Burkhardt, N. Chikuma, M. A. Cortes-Giraldo, S. Elles, V. Grichine, S. Guatelli, et al. Progress in Geant4 electromagnetic physics modelling and validation. Journal of Physics: Conference Series, 664(7):072021, 2015.

\bibitem[Buonomo et al.(2023)]{Buonomo:2023pzi}
B. Buonomo, F. Cardelli, C. Di Giulio, D. Di Giovenale, L. G. Foggetta, C. Taruggi. The Frascati Beam Test Facility. arXiv:2308.03058 [physics.acc-ph], 2023.

\bibitem[Buonomo et al.(2016)]{buonomo2016hardware}
B. Buonomo, C. Di Giulio, L. G. Foggetta, P. Valente. A hardware and software overview on the new BTF transverse profile monitor. In: Proceedings of the 2016 International Beam Instrumentation Conference (IBIC'16), Barcelona, Spain, pp. 11--15, 2016.



\end{thebibliography}
\end{document}